\documentclass[11pt]{article}
\usepackage{graphicx,amsmath}
 \usepackage{amssymb}
\usepackage{lineno,color}
\usepackage{url}

\begin{document}
\begin{center}
{\bf\Large Hunting for QCD Instantons~\footnote{Talk given by M.~G.~Ryskin at the Conference  ``Particle physics at intermediate and high energies'', Protvino, June 2-5.}}\\

 \vspace{1cm}
   
 M.~G.~Ryskin$^{a}$, V.~A.~Khoze$^{b}$\\
 
 \vspace{0.7cm}
{\small  $^a$  NRC Kurchatov Institute - PNPI, Gatchina 188300, Russia\\
 
 $^b$ IPPP, Department of Physics, Durham University, Durham, DH1 3LE, UK\\
}
 \vspace{0.7cm}

 \abstract{
 \noindent  
Instantons are non-perturbative classical solution which describe
 transitions between different QCD vacua.
They have never been observed experimentally.
We consider the signatures of the instanton(sphaleron)
 production events and the main QCD backgrounds.
The possibilities to search for the QCD instantons in the diffractive
(i.e. with a large rapidity gap) events at the LHC and via the spin-spin
correlations between two hyperons at NICA are discussed.  
}   
 
 \vspace{0.5cm}
 
 E-mail: \url{v.a.khoze@durham.ac.uk},\\ \url{ryskin@thd.pnpi.spb.ru}
 
  \end{center}

 \section{Introduction}

Instantons are non-perturbative classical solutions of Euclidean equations of motion in non-abelian gauge theories~\cite{BPST}. 
In the semi-classical limit, instantons provide dominant contributions to the path integral and describe quantum tunnelling between\\ different vacuum sectors of the theory \cite{tH, Callan:1976je, Jackiw:1976pf}. 
QCD instantons are either directly responsible for generating, or at least contributed to, many key aspects of non-perturbative low-energy dynamics of strong interactions~\cite{tHooft:1986ooh, Callan:1977gz, Novikov:1981xi, Shuryak:1982dp, DP, DP1, Schafer:1996wv}. These include the role of instantons in the breaking of the $U(1)_A$ symmetry and the spontaneous breakdown of the chiral symmetry, the formation of quark and gluon condensates, $<0|\bar qq|0>$ and \\$<0|G^a_{\mu\nu}G^a_{\mu\nu}|0>$ and so on.

In QCD, the instanton configuration consists of the gauge field,
\begin{eqnarray}
A_\mu^{a\, {\rm inst}}(x)=\frac{2}{g} \frac{\bar{\eta}^a_{\mu\nu} (x-x_0)_\nu}{((x-x_0)^2+\rho^2)}\,,
\end{eqnarray} 
along with the fermion components for light ($m_f <1/\rho$) fermions,
\begin{equation}
\bar{q}_{Lf} = \psi^{(0)}(x) \,, \quad q_{Rf} = \psi^{(0)}(x) \,.
\end{equation}
The gauge field $A_\mu^{a\, {\rm inst}}$ is the Belavin-Polyakov-Schwartz-Tyupkin (BPST) instanton solution~\cite{BPST}  of the self-duality equations.  Here $\rho$ is the instanton size and $x_0$ is the instanton position. 
Constant group-theoretic coefficients $\bar{\eta}^a_{\mu\nu}$ are the \\'t Hooft $\eta$ symbols defined in \cite{tH}. The fermionic components $\psi^{(0)}$ are the corresponding normalised solutions of the Dirac equation
 $\gamma^\mu D_\mu[A_\mu^{a\, {\rm inst}}] \psi^{(0)} \,=0$. These are the fermion zero modes of the instanton.
The instanton configuration is a local minimum of the Euclidean action, and the action on the instanton is given by
$S_I= \frac{8\pi^2}{g^2} = \frac{2\pi}{\alpha_s}$.

The instanton configuration has topological charge equal to one, and thus, due to the chiral anomaly, the instanton processes violate chirality. 

At large $x\to\infty$ distances, the instanton is a pure gauge field
$$g\frac{\tau^a}2A^a_\mu\to iS\partial_\mu S^+$$
with $S=i\tau^+_\mu x_\mu/\sqrt{x^2}$\\
However, for $x\neq\infty$, it is the real transverse gluon field which describes the transition between two different (in gauge) QCD vacuums.

If the energy is sufficiently large to produce these transverse gluons, it is not a tunnel, but a real transition in which several new gluons are emitted.
In such a case, it is called a sphaleron; however, below we will use the term  instanton.\\

In terms of Feynman diagrams, the sphaleron is described by the multiparticle vertex which emits some number of isotropically distributed gluons plus the  $\bar qq$ pairs of light quarks.

If the instanton is produced by a two-gluon initial state, the final state of this instanton-mediated process will have $N_f$ pairs of quarks and anti-quarks with the same chirality,
\begin{equation}
\label{e2}
g+g\to n_g\times g + \sum^{N_f}_{f=1}(q_{Rf}+\bar q_{Lf})\ ,
\end{equation} 
where $N_f$ is the number of light (relative to the inverse instanton size, $m_f <1/\rho$) flavours.\\

It is quite important to observe the QCD instantons to answer the question whether indeed in QCD we have many different (in gauge) equivalent vacuums. 
If the instanton does not exist, it means that in our theory we have to consider only the fields which decrease at infinity faster than $1/x$ and therefore the instanton solution is forbidden.

Up to now, the instantons/sphalerons have never been observed 
experimentally~\footnote{It was noticed by Bjorken~\cite{Bj} that the large ($\sim 5\%$) branching of 
$\eta_c\to \eta\pi\pi,\ \eta'\pi\pi\, \mbox{and}\, \bar KK\pi$ decays,  naturally produced via the 't~Hooft instanton-induced interaction 
 ${\cal L}\sim (\bar cc)(\bar uu)(\bar dd)(\bar ss)$
  can be considered as the experimental indication of the presence of  QCD instantons (see also~\cite{14,15}).}.

The possibility to observe instantons in inelastic proton-proton collisions at hadron colliders was considered in~\cite{BR, BR1} and more recently in~\cite{KKS, KMS}.

The problem is that a large size instanton which has a sufficiently large production cross section emits rather small number of relatively low energy ($E\sim 1.5/\rho$) minijets and it is challenging to distiguish it from other soft QCD processes while the small size instanton/sphaleron has a very small cross section suppressed by the factor $exp(-S_I)=exp(-2\pi/\alpha_s(\rho))$.

\section{Instanton signatures}
As was explained above, the instanton/sphaleron production looks like the creation of some fireball which emits 
isotropically a large number of minijets $N_{jet}\sim 1/\alpha_s(\rho),\; E_j\sim 1.5/\rho$.

Thus, the main signature of the small size instanton is the
large multiplicity os secondaries and a large transverse energy $E_T=\sum_i |p_{ti}|$ concentrated in a limited rapidity interval.\\

Besides this, we have to expect  additional pairs of strange ($\bar ss$) and charm, $\bar cc$ (if $m_c<1/\rho$) 
quarks. Moreover, the polarization of these extra quark and the antiquark are correlated ($q_R$ and $\bar q_L$). That is, contrary to perturbative QCD ($g\to \bar qq$ or $gg\to \bar q q$), an instanton does not conserve the light quark helicity.

To check that this fireball radiates isotropically, we introduce the sphericity $S$ calculated as
$S=(3/2)(\lambda_2+\lambda_3)$ where $\lambda_i$ are the eigenvalues of $S^{\alpha\beta}$ with
$$ S^{\alpha\beta}=\frac{\sum_{i}p^\alpha_i p^\beta_i}{\sum_i |\bold{p}_i|^2} \ .$$
Note that the trace $\sum_\alpha S^{\alpha\beta} =1$ and we choose $\lambda_1>\lambda_2,\lambda_3$.\\
For an exactly isotropic distribution $\lambda_1=\lambda_2=\lambda_3=1/3$ and $S=1$. On the other hand, the perturbative QCD dijet (back-to-back) production
gives $S$ close to zero.\\

It was shown in \cite{12} that studying at 13 TeV the sphericity distribution of events with the sphaleron mass $M_I=20 - 40$ GeV (actually this is the mass of the tracks observed in the central ($|\eta|<2.5$) region), we should expect about 75\% excess at $S>0.85$ caused by the sphaleron. However, this is within the uncertainty of the model. Recall that the soft background was generated by\\ PYTHIA8 Monte Carlo (MC), and no general-purpose MC was tuned for such exotic events.

To be sure that we observe exactly the instanton,  we have to use at least two signatures. In \cite{12} it was proposed
to select the events with a large number of displaced vertices,
$N_{disp}$. \\These vertices are caused by the\\ charmed secondaries decay, and an additional $\bar cc$ pair is used as the second instanton signature. Indeed, now at $N_{disp}>6$, the instanton ($M_I=20-40$ GeV) signal exceeds the background more than 10 times.  However, after this selection, the cross-section falls by 20-30 times.\\

Recently, the CMS collaboration reported the results of event shape measurement in minimum-bias events
from proton-proton collisions at $\sqrt s=13$ TeV in the special low luminosity run with
approximately one inelastic proton-proton collision per bunch crossing.  They observe a consistent trend in which the unfolded data are more
isotropic than the simulation \cite{cms1}. \\

Another possibility proposed in \cite{2026} is to study the azimuthal angle, $\Delta\phi$ distribution of (relatively) high $E_T$ jets. In perturbative QCD, the hard matrix element produces back-to-back dijets with $\Delta\phi\simeq 180^o$ while from the instanton we may get a much smaller $\Delta\phi$. 

\section{Bachground}

Let us consider the main sources of background which can mimic the instanton signal.

\medskip

The cross-section of instanton production falls steeply with $M_{\rm inst}$ mainly due to the factor\\ $\exp(-S_I)=\exp(-2\pi/\alpha_s(\rho))$ in the amplitude. At  smaller values of $\rho$, the instanton action
$S_I=2\pi/\alpha_s(\rho)$ increases since the QCD coupling $\alpha_s$ decreases. Calculations show that the elementary cross-section of the parton-level subprocess  falls approximately as  
\begin{equation}
\hat\sigma_{\rm inst}\propto M^{-6}_{\rm inst}\,,
\label{e_inst_simp}
\end{equation}
over a broad low-to-intermediate energy range.

The background caused by the purely perturbative QCD subprocess, 
$gg\to N{\rm minijets}$, decreases only logarithmically. 

For the perturbatively formed\\ `hedgehog' configuration of $N$ final state jets we would expect
\begin{equation}
\label{e3}
\sigma_{\rm pQCD}({gg\to N_j})\, \sim\,  \frac{16\pi}{M^2}\alpha^N_s(M)\,,
\end{equation}
where $M$ denotes the invariant energy of the perturbatively formed cluster of minijets.
Thus, at sufficiently large values of $M_{\rm inst}$, the instanton signal~\eqref{e_inst_simp} will become negligible relative to the purely perturbative QCD production~\eqref{e3}. Note that these higher-order subprocesses are not included in today's Monte Carlo generators.

\medskip

In the regime of interest of moderately small $M_{\rm inst}$, we face, however, another problem that needs to be addressed.
The instanton event can be mimicked by multiple parton interactions (MPI). Indeed, the double (triple, ..., $n$) parton scattering produces a few pairs of jets (dijets) which would look like a fireball, thus obscuring the final state signature of the genuine instanton signal.
 At relatively low $M_{\rm inst}$ when the transverse energies of produced jets become small, the MPI processes will dominate.
 
 To suppress the MPI contribution, we can search for the QCD instanton in a large $Q^2$ electroproduction (DIS)~\cite{Moch:1996bs, Ringwald:1998ek} or in the diffractive events with a large rapidity gap (LRG).
 
 Recall that the instantons were \\seached at HERA \cite{Adloff:2002ph, Chekanov:2003ww} but have not been observed. Maybe the future electron-ion collider will be luckier (if the integrated luminosity is sufficiently large).

\section{Instanton in diffractive events}

As discussed above, we expect a large `underlying event' background both in the region of large and small instanton masses. 
It was shown in~\cite{Sas:2021yxx} that the events caused by the MPI manifest high sphericity S. 

However, the low $M_{\rm inst}$ background caused by MPI can be effectively suppressed by selecting events with  Large Rapidity Gaps (LRGs). Indeed, each additional 'parton-parton$\to$ dijet' scattering is accompanied by the colour flow created by the parton cascade needed to form the incoming partons. This colour flow produces secondaries which fill the LRG. The LRG survival probability, $S^2$, (i.e. the probability not to destroy the LRG) is rather small, $S^2\leq 0.1$, see e.g.~\cite{LRG}.
 Thus, the probability of observing $n$ additional branches of parton-parton interactions in LRG events is suppressed by the factor $(S^2)^n$. 

Since the relatively heavy instanton produces  a rather large number of jets, we are looking for 
high multiplicity events which:
\begin{itemize}
\item do not contain high-$E_T$ jets, and 
\item still have a large density of the transverse energy, $\sum_i dE_{Ti}/d\eta\sim M_{\rm inst}/3$ (the sum is over all secondary particles in the given $\eta$ interval).
\end{itemize}
Indeed, the instanton of mass 30 GeV produces about 17 jets (9 gluons plus 4 light $\bar qq$ pairs).
The energy of each jet $E_{Ti}\sim 1.5/\rho\sim 2$ GeV.
After hadronization in such an event, we expect about 40-60 particles. The large multiplicity can be used as the main (or additional) trigger to select the events of interest.

 Thus, we proposed~\cite{hunt} to select events with charged particle multiplicity $N_{ch}>20$  and transverse energy $\sum_i E_{Ti}>15$ GeV within the rapidity interval $0<\eta<2$ (see Fig.1). Moreover, these particles should not have the $p_{Ti}>2$ GeV to exclude the high $E_T$ jets. Expected after this selection, sphericity is shown in Fig.2 (due to a limited rapidity interval, we use the 'transverse' sphericity accounting just for the transverse ($x,y$) components of $p_T$).
 It is seen that after such a selection, the instanton signal
 strongly exceeds the background.
Note that the elementary (parton level) instanton cross-section is predicted to be rather large ($\sim 1 \mu b$), leading after selection to the signal cross-section at $\sqrt s=13$ TeV of about $1 nb$.
\begin{figure} [t]
\begin{center}
\includegraphics[trim=0.0cm 0cm 0cm 0cm,scale=0.35]{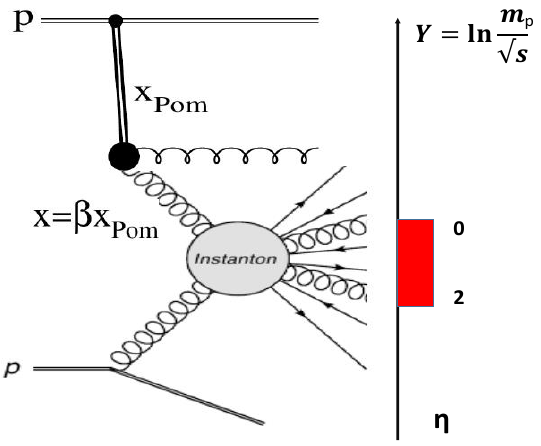}
\caption{\small Instanton production in a diffractive process with an LRG. The pomeron exchange is shown by the thick, doubled line. The red bar shows the considered range of $\eta$, and $Y$ indicates the incoming proton position in rapidity. As shown in the diagram, some secondaries will be produced outside this range but will not be used when calculating $E_{T}$ or $N_{ch}$.}  
\label{fig:1}
\end{center}
\end{figure}

\begin{figure} [t]
\begin{center}
\includegraphics[trim=0.0cm 0cm 0cm 0cm,scale=0.35]{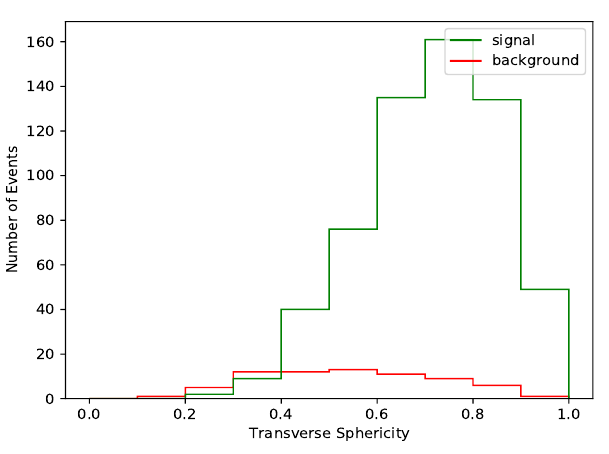}
\caption{\small  Distribution over the transverse sphericity of the charged hadrons produced in the events with the instanton (green) in comparison with the expected background (red). }
\label{fig:2}
\end{center}
\end{figure}

\section{Spin-spin correlations}
As we mentioned before, the sphaleron production is accompanied
by the creation of the light $\bar q_{Lf} q_{Rf}$ pair. If the sphaleron was created by the  quark, the final quark helicities are correlated with the incoming helicity 
$$ q_{Li}+q_{Lk}\to n_gg+\sum_f(q_{Rf}+\bar q_{Lf}),$$
where $i\neq k$ and $f\neq i\; f\neq k$.\\

In comparison with perturbative QCD, we observe the helicity non-conservation~\footnote{This is the QCD analogue of the baryon charge non-conservation in the electroweak instanton case.}.
This can be studied experimentally at NICA, say in 
\begin{equation}
\label{eq:spin}
p_\uparrow+p\to \Sigma+X
\end{equation}

process~\cite{spin}.
First, at the quark level, the instanton doubles the incoming
polarization. Instead of one left $u_L$ quark, it produces two
right quarks – $u_R$ and $s_R$. To distinguish the ’left’ and ’right’ quarks, we need weak interaction, that is, the weak decay of $\Sigma$ or $\Lambda$ hyperons. Since within the SU(6) quark model the $\Sigma$ hyperon contains the vector
 $(ud)$ diquark  there should be a
chance to observe in $\Sigma$ the presence of two right quarks. This
is impossible in the $\Lambda$ case where the $(ud)$ diquark
is the scalar. On the other hand, $\Lambda$ has the advantage – its polarization (in this model) is equal to the polarization of the $s$-quark.

Another possibility is to produce the pair $\bar\Lambda\Lambda$ and to check that the $\Lambda$ is {\em right}-handed while the $\bar\Lambda$ is {\em left}-handed.

Here we deal with the rather large size (low mass) instanton,
and the expected instanton cross section is not too small ($\sim 1 mb$). Accounting for the probability to form the hyperon, we may expect $\sim 1 \mu b$.

To confirm that it was the instanton/sphaleron, we could
observe a larger than usual (at this energy) multiplicity and a stronger energy ($\sqrt s$) dependence and/or the additional spin-
spin correlation, when the second beam is polarized as well, and the
instanton is created via the quark-quark collision (instanton
can absorb two {\em left-handed} quarks only).

\end{document}